\documentstyle[11pt,paspconf,epsf]{article}
\def \bc {\begin{center}}
\def \ec {\end{center}}
\def \be {\begin{equation}}
\def \ee {\end{equation}}
\def \nh {$N_{\rm H}$}

\def \nhgal {$N_{\rm H~I}^{\rm {\tiny Gal}}$}

\begin{document}

\title{The Ionizing Continuum of Quasars}

\author{Ari Laor}
\affil{Physics Department, Technion, Haifa 32000, Israel}

\begin{abstract}
The ionizing continuum shape of quasars is generally not directly observable, but
indirect arguments, based on photoionization models and thin accretion disk 
models suggest that it should peak in the extreme UV, and drop steeply
into the soft X-ray regime. However, recent observations of very soft
X-ray emission in low $z$ quasars, and far UV emission of high $z$ quasars,
suggest that the ionizing continuum of quasars 
does not peak in the extreme UV, and may extend as a single power law
from $\sim 1000$~\AA\ to $\sim 1$~keV. If true, that has interesting
implications for photoionization models and for accretion disk models.
The proposed revised continuum shape will be tested directly in the near 
future with {\em FUSE}. 

\end{abstract}


\keywords{quasars,massive black holes, host galaxies}

\section{Introduction}
What is the shape of the ionizing continuum (1-$\sim 10$ Rydbergs) of
quasars?  This question is interesting because this is where quasars'
continuum emission peaks (e.g. Sanders et al. 1989; Elvis et al. 1994), 
and it therefore provides
an important clue to the nature of the continuum emission mechanism.
It is also interesting because this spectral shape controls the H to
He photoionization ratio, and the heating per ionization of the ionizing
continuum, and is thus important for understanding the physical parameters
of the photoionized gas in quasars.

The best guess mechanism for the continuum emission mechanism in
quasars is accretion of gas into a massive black hole (e.g. Rees 1984). 
The most
detailed models calculated are for thin accretion disks. One can thus use
theoretical accretion disk spectra to predict the ionizing continuum
shape based on the observed optical-UV emission. 
The implied shape tends to peak in the extreme UV (EUV, e.g. Laor 1990). 
One can also constrain the EUV spectral shape using the He~II~$\lambda 1640$
equivalent width. This line is presumably a pure recombination line, and
thus it can be used as a rather accurate measure of the number of He~II ionizing
photons. Mathews \& Ferland (1987) used this argument to deduced an
ionizing continuum which peaks at a few Rydbergs. Thus, it appeared that
both accretion disks and photoionization models indicate that the
ionizing continuum peaks in the EUV. 

However, these arguments are indirect, and one would like to get the best
possible observational constraints on the ionizing spectral shape. 
To address this question we carried out a large program with the
{\em ROSAT} position sensitive proportional counter (PSPC). The results
were surprising, as further described below (for more details see
Laor et al. 1994; 1997).

\section{How to Observe?}

The large Galactic opacity
prevents a direct observation of the EUV in quasars. One alternative
is to observe the UV spectra of very high redshift 
quasars. The other alternative, adopted in our study, is to go to the other 
side of the Galactic opacity
barrier and to observe low redshift quasars in very soft X-rays. 

X-ray observations below 1~keV prior to ROSAT indicated a spectral
steepening, or equivalently an 
excess emission, relative to the flux predicted by an extrapolation of
the hard X-ray power-law (e.g. Arnaud {\it et al.} 1985; 
Wilkes \& Elvis 1987; Turner \& Pounds 1989). In some objects the excess 
could be described as
a very steep and soft component, which is consistent with the Wien tail of a hot
thermal component dominating the UV emission. However, these studies were 
limited by the low signal to noise ratio (S/N) and energy resolution 
of the {\em EINSTEIN} IPC, and the {\em EXOSAT} LE detectors, in 
particular in the 
crucial energy range below 0.5~keV. This prevented an accurate determination 
of the soft X-ray emission spectrum of quasars. Furthermore, the objects studied
do not form a complete sample, and these results are likely to be biased 
by various
selection effects which were not well defined a priori. In particular, most 
studied objects are nearby, intrinsically X-ray bright, AGNs.

The PSPC detector aboard {\em ROSAT} had a significantly improved sensitivity,
energy resolution [$E/\Delta E=2.4(E/1~{\rm keV})^{1/2}$ FWHM], 
and spatial resolution
below $\sim 2$~keV, compared with previous detectors (Tr\"{u}mper 1983). 
We used this detector to make an accurate determination of the soft X-ray 
properties of a well defined, complete, and otherwise well explored, 
sample of quasars.

\section{What to Observe?}

 We found the BQS sample, a subset of the PG survey 
defined 
by Schmidt \& Green (1983), to be particularly suitable for our purpose 
for the following reasons:
1. These objects are selected only by their optical
properties, thus they are not directly biased in terms of
their X-ray properties.
2. This sample has already been studied extensively, and in a uniform 
manner in other parts of the spectrum, including the radio  
(Kellerman et al. 1989; Miller, Rawlings \& Saunders 1993), 
the mid- to far-infrared (Sanders et al. 1989), 
and the near-infrared to optical (Neugebauer et al. 1987). 
High quality optical spectroscopy was obtained by 
Boroson \& Green (1992), {\em HST} FOS spectroscopy was obtained by
Wills et al. (1998, these proceedings), and {\em ASCA} and {\em SAX}
X-ray spectra were obtained by George et al. (1998) and Fiore et al. (1998).  
 These studies of the PG quasars provide us with the most complete and 
coherent picture of the emission properties of bright AGNs, and allow us to 
make a detailed study of possible correlations between the 
soft X-ray properties and various other
emission properties.
3. This sample includes a large fraction of the brightest
known quasars, thus rather high S/N spectra could be obtained 
within a reasonable amount of spacecraft time. 

The complete PG sample includes 114 AGNs, of which 92 are quasars 
(i.e. $M_B< - 23$). 
We selected a subsample of the PG quasars which is optimally suitable 
for soft X-ray observations by the following two selection
criteria: 
1. $z\le 0.400$. This prevents the rest-frame 0.2~keV from being redshifted 
beyond the observable range.  
2. \nhgal$< 1.9\times10^{20}$ cm$^{-2}$, where \nhgal\ is the H~I Galactic 
column density as measured in 21~cm. This low \nhgal\ cutoff is critical for 
minimizing the effects 
of Galactic absorption. This cutoff implies an upper limit on the Galactic
optical depth in our sample of $\tau_{\rm 0.2 keV}< 1.6$ 
(Morrison \& McCammon 1983). Even with this low \nhgal\ cutoff, 
no photons below 0.15~keV can be detected. This is because
the opacity of the Galaxy increases as $\sim E^{-3}$, giving 
$\tau_{\rm 0.1 keV}=N_H/1.77\times10^{19}$, while the effective area of the
PSPC drops rapidly below 0.15 keV. As a result practically no photons 
below 0.15 keV can be detected from the quasars (although the formal lower 
limit of the usable channels on the PSPC is 0.1 keV).
These criteria limited our sample to 23 quasars, which should be 
representative of the low-redshift, optically-selected quasar population. 

Accurate values of \nhgal\ are crucial, even for our low \nhgal\ 
sample, in order to make an accurate determination of the intrinsic
soft X-ray spectrum. The \nhgal\ values were taken from
Elvis, Lockman \& Wilkes (1989), Savage et al.
(1993), Lockman \& Savage (1995), and the recent extensive
measurements by Murphy et al (1996). All these
measurements of \nhgal\ were made with the 140 foot telescope of the NRAO at
Green Bank, WV, using the ``bootstrapping'' stray
radiation correction method described by Lockman, Jahoda, \& McCammon 
(1986), which provides an angular resolution of 21', and an uncertainty of
$\Delta$\nhgal=$1\times 10^{19}$ cm$^{-2}$ (and possibly lower for our
low \nh\ quasars). This uncertainty introduces a flux
error of 10\% at 0.2~keV, 30\% at 0.15~keV, and nearly a factor of 2 at
0.1~keV. Thus, with our accurate \nhgal\ reasonably accurate fluxes can be 
obtained down to $\sim 0.15$~keV. 

\subsection{The Soft X-ray Continuum}

All the objects in our sample were detected with the PSPC, and 
high quality spectra were obtained for most objects.
The spectra of 22 of the 23 quasars are consistent, to within
$\sim 30$\%, with a single power-law model at rest-frame $0.2-2$~keV. 
There is no evidence for significant soft excess emission with 
respect to the best fit power-law. We place a limit (95\% 
confidence) of 
$\sim 5\times 10^{19}$~cm$^{-2}$ on the amount of excess foreground 
absorption by cold gas for most of our quasars. The limits are
$\sim 1\times 10^{19}$~cm$^{-2}$ in the two highest S/N spectra. 

Significant X-ray absorption ($\tau>0.3$) by partially ionized gas 
(``warm absorber'')
in quasars is rather rare, occurring for $\le 5$\% of the population,
which is in sharp contrast to lower luminosity Active Galactic Nuclei 
(AGNs), where significant absorption
probably occurs for $\sim 50$\% of the population.

For the complete sample we find 
$\langle\alpha_{ox}\rangle=-1.55\pm 0.24$, and
$\langle\alpha_x\rangle=-1.62\pm 0.45$. This may appear to suggest
that $\langle\alpha_{ox}\rangle\simeq \langle\alpha_x\rangle$,
as proposed by Brunner {\it et al.} (1992) and Turner, George 
\& Mushotzky (1993). However, this relation 
does not hold when the sample is broken to the RQQ where 
$\langle\alpha_{ox}\rangle=-1.56\pm 0.26$, and 
$\langle\alpha_x\rangle=-1.72\pm 0.41$, 
and to the RLQ where 
$\langle\alpha_{ox}\rangle=-1.51\pm 0.16$, and
$\langle\alpha_x\rangle=-1.15\pm 0.27$.

A significantly flatter $\langle\alpha_{ox}\rangle$ is obtained when the
three X-ray weak quasars (see Laor et al. 1997), and the absorbed quasar 
PG~1114+445, are excluded.
Thus, ``normal'' RQQ quasars in our sample have
$\langle\alpha_{ox}\rangle=-1.48\pm 0.10$, $\langle\alpha_x\rangle=-1.69\pm 
0.27$, while for the RLQ
$\langle\alpha_{ox}\rangle=-1.44\pm 0.12$, $\langle\alpha_x\rangle=-1.22\pm 
0.28$, where the $\pm$ denotes here and above the dispersion about the mean,
rather than the error in the mean. 

\subsection{The EUV Continuum}

Zheng et al. (1997) have constructed a composite quasar spectrum
based on {\rm HST} spectra of 101 quasars at $z>0.33$.
They find a far-UV (FUV) slope (1050-350\AA) of 
$\langle \alpha_{\rm FUV}\rangle = -1.77\pm 0.03$ for RQQs and
$\langle \alpha_{\rm FUV}\rangle = -2.16\pm 0.03$ for RLQs, with
slopes of $\sim -1$ in the 2000-1050~\AA\ regime. The Zheng
et al. mean spectra, presented in Figure 6, together with the PSPC 
mean spectra, suggest that the FUV power-law continuum extends to
the soft X-ray band. In the case of RQQs there is a remarkable agreement 
in both slope and normalization of the soft X-ray and the FUV power-law
continua, which indicates that a single power law continuum component
extends from $\sim 1000$~\AA\ to  $\sim 1-2$~keV.
RLQs are predicted to be weaker than RQQs at $\sim 100$~eV
by both the FUV and the PSPC composites. It thus appears that there is
no extreme UV sharp cutoff in quasars, and no steep soft component 
below 0.2~keV. This implies that the fraction of bolometric luminosity in 
the FUV regime may be significantly smaller than previously assumed.

The UV to X-ray continuum suggested in Figure 1 is very different from the
one predicted by thin accretion disk models and suggested by photoionization
models. In particular, it implies about a four times weaker FUV ionizing
continuum compared with the Mathews \& Ferland continuum that was deduced based 
on the He~II~$\lambda 1640$ recombination line equivalent width.

\begin{figure}
\plotone{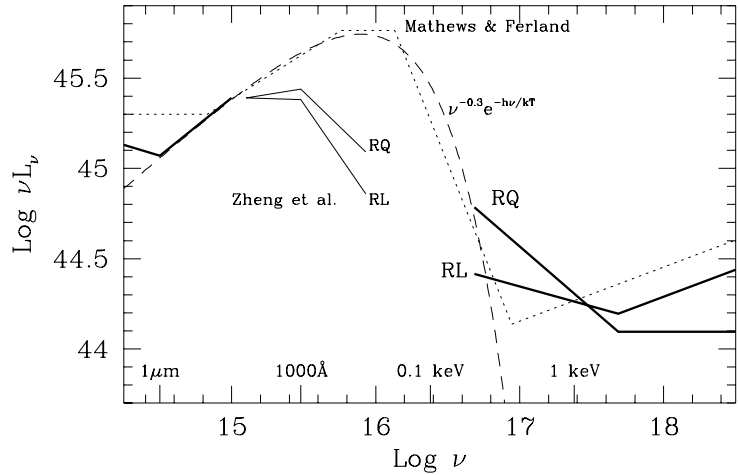}
\caption{Composite optical-soft X-ray spectrum for the RQQs and
RLQs in the Laor et al. sample (thick solid line). Note that despite
the fcat that RLQs are brighter at 2~keV, they are fainter at 0.2~keV.
The Mathews \& Ferland (1987) spectral shape assumes a hard X-ray power
law down to 0.3~keV and a very steep component below 0.3~keV. This
spectral shape is inconsistent with the PSPC results. The Zheng et al.
composites for RLQs and RQQs are plotted in a thin solid line.
They suggest that the FUV power law extends into the soft X-ray regime, 
with no extreme UV spectral break and no steep soft component below
0.2~keV.} \label{fig-1}
\end{figure}

\section{What does it Mean?}

\subsection{Photoionization models}

Korista, Ferland \& Baldwin (1997) discuss possible ways to reconcile
the revised ionizing spectral shape with photoionization models.
They find that there is no way to adjust the BLR parameters to produce the 
observed strength of He~II together with the other UV emission lines, and
so they conclude that either the ionizing continuum is anisotropic, and the
BLR sees a harder continuum than what we see, or that the interpolation
between the FUV and soft X-ray emission is wrong, and there
is an EUV peak near 4 Rydbergs. 

An anisotropic ionizing continuum is naturally produced by thin 
accretion disks, as the radiation from the hottest inner parts of
the disk is deflected towards low inclination angles by the combined
effect of Doppler beaming and gravitational deflection (e.g. fig.8 in 
Laor \&
Netzer 1989). Unified models of AGNs (e.g. Antonucci 1993), as well as 
X-ray spectroscopy of the Fe K$\alpha$ line (Nandra et al. 1997), indicate
that quasars are generally seen not too far from face-on. Thus, the BLR is
most likely spread at high inclination angles, together with the rest of
the obscuring gas, and the observed continuum will always be softer than 
the one incident on the BLR.

The other possibility of an EUV peak may have a physical explanation
as a bound-free He~II emission edge produced in the disk's atmosphere
(Hubeny \& Hubeny 1997, 1998). The problems with this explanation
is that a strong He~II emission edge requires fine tuning of the disk model 
parameters. It also requires fine tuning of the spectral shape so that
the emission shortward of the EUV peak will look like a smooth extension
of the emission longward of the peak.

\subsection{Accretion disk models}

Accretion disks inevitably produce a spectral shape which rises slowly
with frequency and then drops steeply above some cutoff frequency. This just 
reflects the fact that the disk is powered by gravity, and that the dissipated 
energy is radiated locally (and that it has an inner edge). The revised
ionizing continuum drops much more slowly than possible with any form of
simple thin accretion disk models (e.g. Fig.7 in Laor et al. 1997). This 
slow drop
may reflect a drop in the disk radiation efficiency at small radii, either
due to the disk becoming optically thin, so that the viscous infall time
is shorter than the gas cooling time, or if the optical depth remains large, 
the radiative efficiency may drop due to trapping of the outgoing radiation 
in the disk, and its advection beyond 
the black hole event horizon. Alternatively, part of the dissipation may 
occur
in a warm corona above the disk, which will turn the disk exponential tail
EUV emission passing through it into a power law tail 
(e.g. Czerny \& Elvis 1987).

\section{Possible Caveats}

\subsection{Is the PSPC well calibrated at low energy?}

Both {\em EINSTEIN} and {\em ASCA} observations of quasars 
generally suggest a flatter soft X-ray power-law emission.
This raises the possibility that the PSPC may be badly calibrated,
and that the FUV - soft X-ray match may just be a coincidence. 
A proper evaluation of the calibration of these X-ray telescopes
is well beyond the scope of this contribution. It is sufficient
to say that there is no consensus about this issue in the X-ray community.

However, the following result suggests that the PSPC is most likely
well calibrated below 0.5~keV. Figure 2 compares the Galactic 
\nh\ deduced from the accurate 21~cm 
measurements with the best fit X-ray column deduced using \nh\ as a free
parameter (Laor et al. 1997). In most objects the two columns agree to
within $\sim 1 \sigma$. Two objects, PG~1116+215 and PG~1226+023,
have a very high S/N PSPC spectrum, and for these 
\nh(X-ray) is very well determined 
(to within $0.8-1\times 10^{19}$~cm$^{2}$),
yet this column is still consistent with \nh(21~cm), indicating that both 
methods agree to 5-7\%. This remarkably good match implies that the 
PSPC is very unlikely to be have a significantly biased calibration
below $\sim 0.5$~keV, where the ISM absorption becomes significant. 
The \nh(X-ray) vs. \nh(21~cm) match also has various interesting physical
implications, as further discussed in Laor et al. (1997).   

It is also likely that the PSPC is well calibrated above 0.5~keV.
Laor et al. (1994, 1997) fitted each object above 0.5~keV, to look
for spectral curvature by comparing this fit to the fit to the whole
PSPC band (their fit 3). They found that the mean spectral slope 
above 0.5~keV is not significantly different from the mean slope over
the whole PSPC band. This suggests that the PSPC is also well calibrated
above 0.5~keV. Otherwise, the PSPC calibration above 0.5~keV needs to biased
in such a way so as to just compensate for an intrinsic slope change above
0.5~keV.

\begin{figure}
\plotone{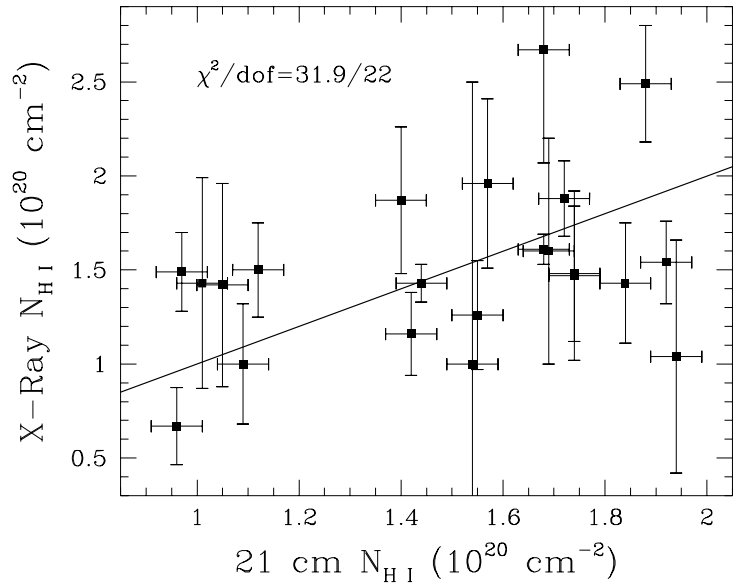}
\caption{The H~I column determined by 21~cm measurements versus the best fit H~I
column determined by a power-law fit to the quasars PSPC spectrum. 
The assumption that both values are equal, indicated by the straight
line, is acceptable at the 8\% level. Note, in particular, the two highest
S/N objects, which deviate from the straight line by less than 
$1\times 10^{19}$~cm$^{-2}$. The agreement between the two measures of
\nh\ indicates a lack of intrinsic cold gas absorption in quasars, and 
that H~I/He~I$\simeq$H/He in the ISM at high Galactic latitudes.
} \label{fig-1}
\end{figure}

\subsection{Is it dust reddening?}
Zheng et al. corrected their individual quasar FOS spectra for Galactic
extinction and also made a statistical correction for absorption by the
Lyman forest. However, no correction was applied for reddening intrinsic
to the quasars. The Galactic extinction curve rises steeply below
2000~\AA\ and relatively small extinction in the optical may induce
significant reddening in the FUV. If quasars have dust with a Galactic 
extinction
curve one may worry that the observed steepening below 1000~\AA\ is induced
by dust. 

The dust opacity for a variaty of grain compositions peaks at 
$\sim 700-800$~\AA\ and drops steeply at shorter
wavelengths, to about 1/3 of the peak opacity at $\sim 300$~\AA\ (see
Fig.6 in Laor \& Draine 1993). Thus, if the observed steepening below 1000~\AA\
was due to dust extinction, then the spectrum at $\lambda<700$~\AA\ should
have flattened back due to the decreasing extinction. Since the observed
composite does not show such a recovery, it is not likely that the 
steepening is due to intrinsic dust absorption, whether it is Galactic
dust or dust of other compositions.

\subsection{Are we comparing apples and oranges?}

The Zheng et al. sample includes only $z>0.33$ quasars, and their composite 
FUV slope is based mostly on $z\ge 1$ quasars, while our sample is limited
to $z\le 0.4$ quasars only. Thus the two samples are practically
disjoint. If the FUV to soft X-ray spectral shape is redshift dependent then
we are not comparing similar objects, and the apparent agreement of the FUV
and soft X-ray composites would be just a coincidence.

One clearly needs to explore the FUV to soft X-ray spectral shape in samples
with similar redshifts. A stronger test is to explore whether
the mean FUV and soft X-ray continua agree in a given sample, and the
strongest test is to explore whether they agree for each object in the 
sample. A large program was recently approved for the {\em FUSE}
mission (PI Anuradha Koratkar) to obtain high quality FUV spectra 
for all our sample of 23 quasars. The spectra will be obtained 
down to the Galactic Lyman limit cutoff, i.e. typical rest frame 
$\lambda\sim 750$~\AA, which is well below the 1000~\AA\ break.
This will allow us to clearly determine for each object whether the
FUV and soft X-ray continua agree.

\section{The EUV in Other Types of AGNs}

Seyfert galaxies show flatter $\alpha_{ox}$ than quasars, but they 
also show a flatter $\alpha_x$ than in quasars. This raises the
possibility that these objects may also have a single power law component
extending from the FUV to $\sim 1$~keV, as was suggested by Zheng
et al. (1995) for Mrk 335. {\em FUSE} observations of a large sample
of Seyfert galaxies having high quality PSPC spectra are required
in order to address this possibility, although given the 
low $z$ of Seyferts, it will be possible to probe their continuum
slope only down to rest frame $\lambda 850$~\AA.

BALQSOs appear to have a steep FUV slope (Korista et al. 1992; 
Arav et al. 1998), and they also generally show a steep $\alpha_{ox}$
(Green \& Mathur 1996), most likely due to strong X-ray absorption
(Mathur, Elvis, \& Singh 1995).
Again, this raises the possibility that their FUV extrapolates to the
soft X-ray flux level. {\em HST} and {\em FUSE} observations of a larger sample
of $z\sim 1-2$ BALQSOs can address this possibility. If the FUV of BALQSOs
is indeed generally very steep that would exacerbate the energy budget
problem, which is already significant in normal quasars. Since the 
weakness of the soft X-ray emission is most likely due to absorption,
the steep FUV spectra may also be due to (a wavelength dependent) absorption.

Some AGNs must clearly have an EUV continuum which is very different from
a simple power law. In particular, Puchnarewicz et al. (1994, 1995a, b) 
find
a number of AGNs with an extremely strong soft X-ray component, much above
the UV component. However, these AGNs were selected from the most luminous
soft X-ray sources known, and are thus most likely extreme cases. Other
AGNs have a rather steep UV continuum, but flat $\alpha_{ox}$
(Puchnarewicz \& Mason 1998), possibly due to extinction
of the optical continuum and absorption of the soft X-ray continuum.

\end{document}